\documentclass[aps,prb,floats,twocolumn,showpacs]{revtex4}
\usepackage{graphicx}
\usepackage{dcolumn}
\usepackage{bm}

\begin{document}

\title{Quantitative analysis of the first-principles effective-Hamiltonian
approach to ferroelectric perovskites}

\author{Silvia Tinte, Jorge \'I\~niguez, Karin M. Rabe, and David Vanderbilt}

\address{Department of Physics and Astronomy, Rutgers University,
Piscataway, New Jersey 08854-8019, USA}

\date{\today}

\begin{abstract}

The various approximations used in the construction of a
first-principles effective Hamiltonian for BaTiO$_3$, and their
effects on the calculated transition temperatures, are discussed.  An
effective Hamiltonian for BaTiO$_3$ is constructed not from
first-principles calculations, but from the structural energetics of
an atomistic shell model for BaTiO$_3$ of Tinte {\em et al.}  This
allows the elimination of certain uncontrolled approximations that
arise in the comparison of first-principles effective Hamiltonian
results with experimental values and the quantification of errors
associated with the selection of the effective Hamiltonian subspace
and subsequent projection.  The discrepancies in transition
temperatures computed in classical simulations for this effective
Hamiltonian and for the atomistic shell model are shown to be
associated primarily with a poor description of the thermal expansion
in the former case. This leads to specific proposals for refinements to
the first-principles effective Hamiltonian method. Our results
suggest that there are at least two significant sources of error in the
effective-Hamiltonian treatment of BaTiO$_3$ in the literature,
i.e., the improper treatment of thermal expansion, and the 
errors inherent in the first-principles approach itself.

\end{abstract}


\maketitle

\section{INTRODUCTION}

First-principles methods constitute a powerful tool for the study of
ferroelectric systems.~\cite{van97}
Ground state structures, phonons, spontaneous polarization, and
related properties, including piezoelectric and dielectric tensors,
have been accurately calculated for a wide variety of perovskite oxides
as well as other ferroelectric compounds.

Despite advances in algorithms and computer hardware,
the direct calculation of finite temperature behavior, particularly
phase transitions, is still far beyond reach, as such calculations involve
thousands of atoms.
However, indirect methods have been developed and applied to
a large number of systems, including
BaTiO$_3$,~\cite{zho94,zho95a,tin99} PbTiO$_3$,~\cite{wag97} 
and KNbO$_3$,~\cite{sep97,kra99} and even solid solutions like
Pb(Zr$_{x}$Ti$_{1-x}$)O$_3$,~\cite{bel00,leu02}
Pb(Sc$_{0.5}$Nb$_{0.5}$)O$_3$,~\cite{hem01} and
K(Nb$_x$Ta$_{1-x}$)O$_3$.~\cite{sep02}
In Refs.~\onlinecite{tin99}, \onlinecite{sep97}, and \onlinecite{sep02},
interatomic ``shell-model" potentials were
parametrized by fitting to first-principles results, and
finite-temperature behavior studied by direct simulation of
atomistic systems with forces and energies obtained from these
potentials.  The results in Refs.~\onlinecite{zho94,zho95a,wag97},
and \onlinecite{kra99,bel00,leu02,hem01}
were obtained by an effective Hamiltonian construction
in which the full system is mapped by a subspace projection onto
a statistical mechanical model, with parameters determined from
first-principles calculations of total energies for small distortions
of an ideal crystal with the cubic perovskite structure.
The simple form of the resulting effective Hamiltonian
allows very-large-scale simulations and aids in the
conceptual interpretation of the results.
The two approaches have achieved comparable success in reproducing
many essential features of
the phase transitions of ferroelectrics.
For example, for BaTiO$_3$,~\cite{zho94,zho95a,tin99} the experimentally
observed cubic--tetragonal--orthorhombic--rhombohedral phase sequence
is correctly reproduced.
However, while the experimental values of the transition temperatures
are 403, 278, and 183~K, classical Monte-Carlo simulations of the
first-principles effective Hamiltonian give
300, 230, and 200~K, while the shell-model results are 210, 135 and 100K,
respectively. In both cases a correction for the
local-density approximation (LDA) lattice constant underestimate was included
in the model.

Understanding the origin of this discrepancy with experiment
may help in the development of improved theoretical methods
for the calculation of finite temperature behavior.
Here, we focus our attention on the first-principles
effective Hamiltonian approach.
The discrepancies in the transition temperatures could result from
separate errors introduced at various steps of the
analysis.
We correspondingly classify the errors into five types.
Errors in the configuration energies obtained from first-principles
calculations
will be designated Type~I.
These generally can be systematically reduced, with the exception of the
uncontrolled approximation in the exchange-correlation functional
required for the practical implementation of density functional theory
(``LDA error'').
Type II errors result from the identification of the relevant
degrees of freedom and the projection and approximate
representation of the effective Hamiltonian in the corresponding subspace,
and will be the main focus of our investigation.
Errors in the statistical-mechanical
simulations (Type~III) include finite-size effects and sampling errors.
In the effective Hamiltonian studies to date, it has been feasible to
make these errors relatively negligible.
The importance of Type IV errors, resulting from the classical treatment
of the ions neglecting
quantum fluctuations, has been highlighted in a recent study of
BaTiO$_3$ by \'I\~niguez and
Vanderbilt.~\cite{ini02} The results of this study indicate 
that the
classical approximation {\it raises} the transition temperatures. Thus,
this is not the origin of the underestimate for BaTiO$_3$. 
In fact, a correct, fully quantum-mechanical treatment would 
increase, not decrease,
the transition-temperature discrepancy.
Finally, we note that the experimental samples, even in thermodynamic
equilibrium, contain defects and local
nonstoichiometry which lead to deviations of observed
properties from those of the assumed ideal crystals (Type~V errors).
These crystal imperfections can have various effects on the transition
temperatures that
are in general difficult to model.

To separate and quantify the role of the various errors in producing
the observed discrepancies in transition temperatures, several
different approaches could be applied.
The analysis of Type III and Type IV errors has been discussed
in the previous paragraph.
One way to investigate Type I errors would be completely to redo
the effective Hamiltonian study replacing the LDA
with a generalized-gradient approximation (GGA).
However, the latter has not been found to give systematic improvement in
the overall agreement of calculated properties with experiment,~\cite{gga}
and thus the value of such a labor-intensive investigation is unclear.
In principle, Type II errors could be eliminated by comparing the
effective-Hamiltonian transition temperatures with those obtained
in a fully {\it ab-initio} molecular dynamics or Monte Carlo calculation.
However, as noted above, doing this type of direct calculation
for sufficiently large systems is so computationally demanding that
it is impossible in practice even for benchmarking purposes,
and calculations for small supercells and with reduced
sampling would introduce significant finite-size and statistical errors.

In this paper we develop and carry out an alternative method of isolating
and quantifying
Type II errors, allowing us to discuss possible refinements of
the effective Hamiltonian method to reduce or eliminate them.
We use the total energies computed with the BaTiO$_3$ ``shell model"
interatomic potential of Tinte {\it et al.} \cite{tin99} to construct
an effective Hamiltonian, and compare the computed transition
temperatures with those obtained in direct classical simulations for
the ``shell model" system.
In this comparison, we completely eliminate errors of Types
I, IV and V, and can easily make errors of Type III
negligible. Thus, we can attribute any discrepancies directly
to errors of Type II. While such errors will not be quantitatively
identical to the corresponding errors made in the construction of
the effective Hamiltonian directly from {\it ab-initio} results, the general
accuracy and physical faithfulness of the shell model interatomic
potential to BaTiO$_3$ should render conclusions based on this
analysis quite meaningful.

The paper is organized as follows. Section~II provides technical
details of the BaTiO$_3$ shell-model interatomic potential of Tinte
{\it et al.} that serves as our reference system.  In Sec.~III we
describe
the construction of 
the effective Hamiltonian 
and the parameters obtained by fitting to the shell model, paying special
attention to the approximations and technicalities involved.
In Section~IV we present the results obtained from the
effective-Hamiltonian and shell-model classical statistical-mechanical
simulations.
The discrepancies are analyzed in Section~V, and
possible improvements on the various effective-Hamiltonian
approximations are discussed. Section~VI is devoted to the specific
issue of modeling the thermal expansion within the
effective-Hamiltonian approach. Finally, in Section~VII we present
a discussion of the broader implications of our analysis; in particular,
we speculate on the
relative importance of errors of Types I, II, and~IV in the first-principles
effective-Hamiltonian treatments currently in the literature.

\section{Shell-model interatomic potential}

Of the various types of interatomic potentials, shell models are
uniquely well suited to giving a good description of the lattice dynamics of
perovskite oxides.  The form of the shell-model potential developed for
BaTiO$_3$ in Ref.~\onlinecite{tin99} incorporates earlier observations
that the oxygen shell-core interaction should be nonlinear and
anisotropic.\cite{bilz,lewiscatlow}
Each ion (Ba, Ti or O) is modeled as a massive core linked to a massless 
shell. The core-shell interactions for Ba and Ti are harmonic and isotropic.
An anisotropic core-shell interaction is considered at the O$^{-2}$ ions, 
with a fourth-order core-shell interaction along the O-Ti bond.
In addition to the Coulomb interactions between ion cores and shells, 
the model contains pairwise short-range inter-shell potentials of the
Buckingham type, i.e., $V(r) = a\,\exp(-r/\rho) - c/r^6$.
The Born-Mayer form ($c=0$) is sufficient for the Ti-O and Ba-O
short-range potential, while for the O-O potential the value of $c$ is 
nonzero.
The physically important nonlinearities of the interatomic interactions are
thus naturally incorporated into the form of the potential. 

The material-specific parameters in the interatomic potential were determined 
by adjusting them to fit selected first-principles
results computed using the linearized augmented planewave (LAPW) method. 
It should be noted, however, that the equilibrium lattice constant of the cubic phase
is fitted to the experimental cubic-phase lattice constant extrapolated to 0 K
(3.995 \AA), not the LAPW lattice constant. 
The double wells for polar distortions along (001), (011), and (111) are
satisfactorily reproduced, as are the phonon dispersion curves for the cubic
structure at the experimental lattice constant.
The bulk modulus of the cubic phase and the anomalous Born effective charges are
also
in good agreement with the first-principles results.  
Reference~\onlinecite{tin99} contains further details about 
the construction of the interatomic potential and values of the parameters.

Finite-temperature properties of the system described by this interatomic potential 
are investigated by constant-pressure molecular dynamics (MD) simulations using 
the DL-POLY package,~\cite{dlpoly}
where the adiabatic dynamics of the electronic shells are approximated 
by assigning small masses to them.
A Hoover constant-($\bar{\sigma}$,T)
algorithm with external stress set to zero is employed; all cell lengths and cell angles are allowed to fluctuate.
The time step is 0.4 fs and the total time of each simulation,
after 2 ps of thermalization, is 20 ps.
Results for a 
7$\times$7$\times$7 periodic supercell (1715 ions plus 1715 shells which are additional degrees of freedom) 
were reported in Ref.~\onlinecite{tin99}.
It was shown that the cubic--tetragonal--orthorhombic--rhombohedral
phase sequence is correctly reproduced.
Good agreement with experimental data was obtained for
the structural parameters in the various phases as well as the volume thermal expansion coefficient, showing that 
the most important nonlinearities have been included in the model.
However, the transition temperatures are rather low compared to experiment
(190, 120, and 90~K).
This discrepancy does not affect the present analysis of Type II errors.
In the present work, we have expanded the supercell to 
10$\times$10$\times$10 primitive cells (10000 degrees of freedom).
This yields essentially the same results, except that the
calculated transition temperatures increase slightly (210, 135, and
100~K). Additional MD simulations were performed at constant volume,
using a modified Hoover constant-($\bar{\sigma}$,T) algorithm that 
allows for fluctuations in the cell shape.

\section{Construction of the effective Hamiltonian} 

In this Section we describe the effective Hamiltonian that we have
constructed using the shell model for BaTiO$_3$ of Tinte {\it et
al.}\cite{tin99} as our target system.  The form of the effective
Hamiltonian is identical to that proposed by Zhong
{\it et al.},\cite{zho95a}
except that the
inhomogeneous strain variables found to be unimportant in that study are
not included here.
 
An effective Hamiltonian is a Taylor expansion of the energy surface
of the system around a high-symmetry phase in terms of a set of
relevant degrees of freedom. For ferroelectric perovskites, the most
convenient reference structure is the cubic paraelectric phase. The
relevant degrees of freedom can be identified by studying the energy
changes induced by small (harmonic) perturbations of the reference
structure. The low-energy, typically unstable, distortions are the
relevant ones, and are expressed in the
form of local modes or lattice Wannier functions.~\cite{rab95}
The relevant local modes are those that add up to produce the distorted
ferroelectric ground-state structure.  Also, we
take the homogeneous strains as relevant and include them in the
Hamiltonian.
 
There are two possible ways of performing the harmonic analysis that 
leads to the identification and calculation of the relevant lattice 
Wannier functions. One can study either the force-constant 
matrix (the matrix of second derivatives of the energy with respect
to atomic displacements) or the 
corresponding dynamical matrix. While the former choice leads to a 
better description of the lowest-energy states of the system, the 
latter provides a kinetic decoupling between the relevant and 
{\it irrelevant} (i.e., not considered in the Hamiltonian) degrees of 
freedom. Here we have worked with the force-constant matrix, which is 
more appropriate for the study of equilibrium properties. In any case, 
we find numerically that the force-constant and dynamical matrix 
descriptions are essentially identical.

\begin{table} 
\caption{Expansion parameters of the effective Hamiltonian fitted
to the shell-model BaTiO$_3$ target system.
The notation is taken from Ref.~\protect\onlinecite{zho95a}. All the 
parameters are in atomic units. } 
\begin{ruledtabular}
\begin{tabular}{l|cr|cr|cr} 
On-site & $\kappa_2$ & 0.0562 & $\alpha$ & 0.805 & $\gamma$ & $-$0.849 
        \\\hline 
        &  $j_1$ & $-$0.01424 & $j_2$ & $-$0.01506  \\ 
Intersite & $j_3$ & 0.00422 & $j_4$ & $-$0.00240 & $j_5$ & 0.01956 \\ 
           & $j_6$ & 0.00100 & $j_7$ & 0.00050 \\\hline 
Elastic & $B_{11}$ & 5.42 & $B_{12}$ & 2.06 & $B_{44} $ & 2.07 
        \\\hline 
Coupling & $B_{1xx}$ & $-$4.16 & $B_{1yy}$ & $-$1.19 & $B_{4yz} $ & 
        $-$0.44 
  \\\hline 
Dipole & $Z^*$ & 8.153 & $\epsilon_{\infty}$ &5.24  
\end{tabular} 
\end{ruledtabular}
\label{tab:param}
\end{table} 
 
Once a relevant set of phonon branches~\cite{fn:phonon} has been 
identified, the calculation of the corresponding lattice Wannier 
functions can be done at different levels of approximation.  At the 
crudest level, they can be constructed from phonons at a single $k$ 
point,~\cite{zho95a} more sophisticated schemes allowing for better 
descriptions of the relevant phonons throughout the Brillouin 
zone.~\cite{rab95,ini00}  
Here, we calculate the local modes 
from the unstable phonons at $\Gamma$ (these generate the ferroelectric
structure)
choosing the local modes to be
centered on the Ti atom. The resulting local modes reproduce 
the unstable phonons at zone-boundary points M and R with an accuracy 
above 97\%. We thus do not expect a better local-mode definition to 
constitute a significant improvement of our effective Hamiltonian of 
shell-model BaTiO$_3$.
 
Let $u_{i\alpha}$ denote the local mode amplitude in unit cell $i$ along 
Cartesian direction $\alpha$. Let $\eta_l$ denote the strains, where 
$l$ is a Voigt index. Our effective Hamiltonian can then be written as 
\begin{eqnarray} 
&H_{\rm eff} &=\; E^{\rm self}[u_{i\alpha}^2; u_{i\alpha}^2 
u_{i\beta}^2] + E^{\rm dpl}[u_{i\alpha}u_{j\beta}]+ \nonumber \\ 
&&E^{\rm short}[u_{i\alpha}u_{j\beta}] + E^{\rm elas}[\eta_l \eta_m] 
+ E^{\rm int}[\eta_l u_{i\alpha} u_{i\beta}]. 
\label{eq:heff} 
\end{eqnarray} 
Following Ref.~\onlinecite{zho95a}, we have written the effective Hamiltonian 
as the sum of four terms: 
the on-site self-energy of the local modes $E^{\rm 
self}$, the long-range dipole-dipole interactions between local modes 
$E^{\rm dpl}$, the short-range interactions between local modes 
$E^{\rm short}$, the elastic energy $E^{\rm elas}$, and the 
interaction between local modes and strains $E^{\rm int}$. The 
dependence of each term on the model variables is indicated in 
Eq.~(\ref{eq:heff}). It should be noted that the form of the 
Hamiltonian is greatly simplified by the cubic symmetry of the 
reference structure. For example, there are no odd terms in $u_{i\alpha}$. 
 
The relevant phonon branches are described by the harmonic terms in 
$E^{\rm self}$, $E^{\rm dpl}$, and $E^{\rm short}$. Anharmonic terms 
for the local modes, required to stabilize the low-symmetry phases, 
are included only in $E^{\rm self}$ (the ``local-anharmonicity approximation").
In both $E^{\rm elas}$ and $E^{\rm 
int}$ only the lowest-order terms in the expansion are 
considered. This constitutes the minimal microscopic model for the 
description of ferroelectricity in BaTiO$_3$.  
Including higher-order terms in the Hamiltonian would constitute a systematic 
improvement of the model.

\begin{figure}
\begin{center}
\includegraphics[width=8.6cm,angle=0]{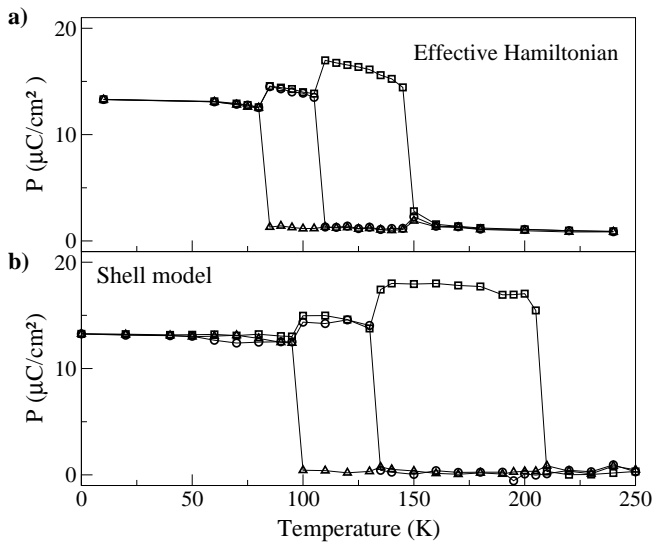}
\end{center}
\caption{Behavior of three Cartesian components of the mean (super-cell
averaged) polarization of BaTiO$_3$ as obtained from (a) MC simulations
using the effective Hamiltonian, and
(b) MD simulations directly from the shell model.}
\label{fig:pol}
\end{figure}

As already mentioned, the Hamiltonian we have just described is 
essentially that of Ref.~\onlinecite{zho95a}, except
the parameters were fitted to a series of shell-model calculations
of total energies and forces, instead of {\it ab-initio}
results, for a set of distorted structures.
Following the notation of Ref.~\onlinecite{zho95a},
we list in Table~\ref{tab:param} the parameters defining our effective 
Hamiltonian for shell-model BaTiO$_3$. 

Monte Carlo (MC) simulations were performed to calculate the
finite-temperature properties of the Hamiltonian.  We simulated a
10$\times$10$\times$10 supercell with periodic boundary conditions,
and typically did 30,000~MC sweeps to equilibrate the system and
50,000 sweeps more to obtain averages of local-mode variables with a
statistical error below $10\%$.  The temperature was increased in
small steps of 5~K.  We monitored the behavior of the homogeneous
strain and the vector order parameter to identify the transitions. The
average local-mode vector is proportional to the polarization.  Note
that, unless it is indicated, the MC simulations for the present
effective Hamiltonian are performed at zero external pressure.

\section{Finite-temperature results}

Figure~\ref{fig:pol}a shows the three components of the
mean polarization as a function of temperature as obtained from
the MC simulations of the effective Hamiltonian,
while Fig.~\ref{fig:pol}b shows the corresponding results
obtained directly from the shell-model MD simulations.
It is apparent that the effective Hamiltonian correctly reproduces 
the sequence of transitions (cubic to tetragonal to orthorhombic
to rhombohedral with decreasing $T$).
However, the agreement is far from perfect for the $T_{\rm c}$
values, listed in Table~\ref{tab:tc}.  Clearly the effective Hamiltonian
underestimates the $T_{\rm c}$'s, especially for the C--T transition
where the transition temperature is too low by $\sim$30\%.

This shows that for an effective Hamiltonian of this form, Type
II errors are quite significant, and in fact are comparable to
the discrepancies
found when comparing BaTiO$_3$
{\it ab-initio} effective-Hamiltonian results against experimental
measurements on real BaTiO$_3$ samples.
This strongly suggests that errors in first-principles calculations
account for at most only part of the latter discrepancy. 
In the following, we will investigate the Type II errors in more detail
and identify approaches to systematic reduction of these errors,
returning to the discussion of first-principles effective Hamiltonians
in Sec. VII.

\begin{table}
\caption {BaTiO$_3$ transition temperatures, in K, between cubic (C),
tetragonal (T), orthorhombic (O), and rhombohedral (R) phases, as
obtained from the shell model and from the effective Hamiltonian.  The last
three rows correspond to effective Hamiltonians modified as indicated
in the text. In the first column, percentage error relative to the shell
model is given in parentheses.}
\begin{ruledtabular}
\begin{tabular}{lccc}
             &  C--T    &    T--O   &    O--R      \\
\hline
Shell model   &  210    &    135   &    100   \\  
Effective Hamiltonian
              &  150 ($-$28\%)   &    110   &   85   \\ \\
{\it H}$_{{\rm eff}}$ different $\alpha$ and $\gamma$
              & 150 ($-$28\%)   &    120   &    100 \\
{\it H}$_{{\rm eff}}$ + {\it p}$_{{\rm eff}}$ `by hand'
              &    185 ($-$12\%)   &    125   &   95 \\
{\it H}$_{{\rm eff}}$ + computed {\it p}$_{{\rm eff}}$
              &   165 ($-$21\%)     &    120   &    90  
\label{tab:tc}
\end{tabular}
\end{ruledtabular}
\end{table}

\section{Analysis of sources of the discrepancies}

In this Section, we analyze three sources of error in the
construction of the effective Hamiltonian that could possibly lead to the
calculated underestimates of the $T_{\rm c}$ values.
First, we focus on the specification of the ferroelectric mode unit 
vector, which determines the precise set of degrees of
freedom described by the effective Hamiltonian. Second, we
consider the effect of the truncation of the Taylor expansion
in the specified degrees of freedom, with particular attention to
the neglect of certain higher-order couplings within the effective 
Hamiltonian subspace.
Third, we consider the consequences of omitting the higher-frequency
phonon branches from the effective Hamiltonian.

\subsection{Specification of ferroelectric local mode vector}

One of the first choices that was made in the construction of the
effective Hamiltonian was the detailed specification of the ferroelectric
local mode vector.  As explained in Sec.~III, we chose a Ti-centered
displacement pattern selected in such a way that a uniform
superposition of these local mode displacements gives a periodic
displacement pattern corresponding to the unstable (ferroelectric) 
mode eigenvector
of the force-constant matrix in the cubic structure.
This is only one of many possible approaches, and questions
may arise as to whether this is the best choice and how much
difference it would make if we had made a different choice.

One way of addressing these questions is to test how completely the
chosen local mode vectors span the space of distortions that are
actually encountered in the full atomistic finite-temperature
shell-model simulation.  We
projected the shell-model MD trajectories at a given temperature
onto the 12 optical zone-center normal modes (i.e., force-constant
eigenvectors) of the cubic phase (four sets of three-fold
degenerate modes).  As expected, we found that the mode branches
included in our effective
Hamiltonian subspace account for almost all ($\sim$90\%) of the observed
atomic displacements.  This suggests that the approximation of
keeping only these modes in the effective Hamiltonian is
a good one.

A second approach is to try a different procedure for defining the
identity of the local mode vector. In particular, one could think of a
construction designed to optimize the description in
the neighborhood of the ferroelectric ground state. For instance, the
local mode vector could be fitted to the ground state of the system
that is obtained when the atomic positions are fully relaxed. Such a
procedure would effectively incorporate the effect of the anharmonic
couplings between included and excluded modes while not increasing the number
of variables considered in the model. In order to quantify the effect
of this change, we assume that this alternative local
mode definition mainly affects the anharmonic parameters
in
\begin{equation}
E^{\rm self} = \kappa_2 u_{i}^{2} 
+ \alpha u_{i}^{4} +  \gamma (u_{ix}^{2}u_{iy}^{2}
+ u_{iy}^{2}u_{iz}^{2} + u_{ix}^{2}u_{iz}^{2}).
\label{eq:eself}
\end{equation}
We thus recalculated $\alpha$ and $\gamma$ exactly to reproduce the
energy of the fully relaxed tetragonal energy minimum, and to get the
best compromise for the energies of the orthorhombic and rhombohedral
minima. By ``fully relaxed'' we mean that the atomic positions were
allowed to relax with the cell constrained to be the equilibrium cubic cell;
this is consistent with the fact that we did not recalculate any
mode-strain coupling parameter. The new $\alpha$ and $\gamma$ are
0.811 and $-$0.916\,a.u.\ respectively, which are very similar to
the values in Table~\ref{tab:param}. The smallness of the correction
reflects the fact that the fully relaxed energy minima are very close
to those described by the original effective Hamiltonian, the
differences being of the order of 0.01~mHa. Keeping all other
parameters unchanged, we repeated the MC simulations at finite
temperature and found that the transition temperatures of the T--O and
O--R transitions (see Table~\ref{tab:tc}) are sensitive
to these small changes in parameter values, giving a 10\% improvement
compared with our original effective Hamiltonian.  However, the large
discrepancy in the C--T transition temperature is unchanged.

We therefore conclude that a change in the definition of the local-mode
displacement pattern is unlikely to be sufficient to eliminate the
discrepancy between the effective-Hamiltonian and shell-model results.
It is necessary, therefore, to look elsewhere.  Nevertheless, the
results do show that the details of the fitting procedure can have a
significant effect on the transition temperatures.

\subsection{Neglect of higher-order terms in the Taylor expansion} 

We now return to our initial choice of relevant degrees of freedom,
and ask whether the corresponding energy landscape is sufficiently
well described by the truncated Taylor expansion that defines the
effective Hamiltonian. The quadratic elastic energy $E^{\rm elas}$ is
easily checked to be accurate. The dipole-dipole interactions in
$E^{\rm dpl}$ will be harmonic as long as the local polarization
is linear in the atomic displacements, and this approximation is valid
for BaTiO$_3$. The terms that 
require further consideration are $E^{\rm self}$, $E^{\rm short}$, and $E^{\rm
int}$.
 
Higher-order terms in $E^{\rm self}$ should aim at a better
description of the double-well potentials associated with the
ferroelectric instabilities. We checked, however, that including
sixth- and eighth-order terms does not improve the fit
significantly. In particular, the well depths, which are the
effective-Hamiltonian feature most directly related to the value of
the transition temperatures, are very well reproduced by the quartic
$E^{\rm self}$. A more accurate description would yield energy wells
around 1\% shallower, which would probably lead to a very tiny
decrease in the C--T transition temperature.

Higher-order terms in $E^{\rm short}$ represent anharmonic couplings
between neighboring local modes and would provide a correction to the
local-anharmonicity approximation. One can fit such terms by looking
at the double-well potentials associated with the antiferroelectric
instabilities of shell-model BaTiO$_3$ at zone-boundary points X and M. The
fourth-order terms associated with such wells will be a combination of
the parameters $\alpha$ and $\gamma$ in Eq.~\ref{eq:eself} and the new
quartic parameters in $E^{\rm short}$. However, we find that these new
quartic parameters are very small and can be safely neglected. More
precisely, they constitute 0.05\% and 5\% of the total fourth-order
term for the X and M instabilities, respectively, and result in
slightly-deeper zone-boundary energy wells. Their probable effect is a
minor decrease in the transition temperatures, because of an enhanced
competition between zone-center and zone-boundary instabilities.

One could think of improving on $E^{\rm short}$ by including couplings
between further neighbors (following Ref.~\onlinecite{zho95a}, we
included couplings up to third neighbors in our Hamiltonian). This
would allow us to improve the description of the dispersion branches
of the relevant modes throughout the Brillouin zone. However, we checked
that if our Hamiltonian is fitted including couplings only up to
second neighbors the transition temperatures change by less than
10~K. Hence, we can assume our Hamiltonian is well converged in this
respect.

Finally, the description of the interaction between strain and local
modes can be improved by including more terms in $E^{\rm int}$.  In
particular, we have found that the $\eta_1 u_{ix}^4$ term is not
negligible and would modify the effective Hamiltonian so as to
yield higher transition temperatures. Specifically, we find that the
coefficient of the $\eta_1 u_{ix}^4$ term is negative and would thus
favor a state in which the system polarizes and expands along a
Cartesian direction. However, in the next section we will see 
that the main Type II error has a different origin, and so we
leave explicit consideration of this correction for future work.

\subsection{Effect of excluded modes: thermal expansion}

Finally, we consider the original decision to
reduce the number of degrees of freedom in the effective Hamiltonian
to one vector degree of freedom per cell to represent ferroelectric
distortions.
Even if an optimal set of local-mode variables is chosen (Sec.~V\,A)
and all necessary terms in the Taylor expansion are kept (Sec.~V\,B),
there still may be errors associated with this fundamental
approximation.  For example, anharmonic couplings between included and
excluded modes are neglected, as are anharmonic couplings between
excluded modes and of excluded modes to strain.

The effects of neglecting these anharmonic couplings are clearly seen
in the thermal expansion.  In fact, in raising
the temperature from 0 to 300\,K in our simulations, we
find that the volume of the shell-model system {\it increases}
by 0.4\%, while that of the effective-Hamiltonian system
{\it decreases} by 0.6\%.
This indicates that the effective Hamiltonian treatment of the
thermal expansion is qualitatively incorrect.  Moreover, given
the well-known sensitivities of the transition temperatures
to volume, this effect could be substantial.
Moreover, it correctly predicts that we would underestimate
transition temperatures, since they are reduced at smaller
lattice constants.

To check whether the thermal expansion effect is responsible for
the dominant errors in $T_{\rm c}$, we made the following test.  We
completely eliminated the volume effect by carrying out both simulations
at a {\it fixed volume} of (4.012\,\AA)$^3$. Using the shell model, we
have found $T_{\rm c}$ values of 190, 130, and 95\,K, while the
corresponding effective-Hamiltonian values are 180, 125, and 100\,K,
respectively. The error in the C--T transition temperature, 
which was around 60\,K
in the zero-pressure simulations, is reduced to $\sim$10\,K.

\subsection{Summary}

We thus arrive at the important conclusion that {\it the poor
description of thermal expansion effects is the dominant source
of error} in the effective-Hamiltonian description.  These
shortcomings in the description of thermal expansion,
and some preliminary attempts to correct for
them, will be described in the following Section.
Smaller errors (probably amounting to no more than 5-10\% of the
$T_{\rm c}$ values) are associated with the other approximations
discussed in Subsecs.~V\,A-B.

\section{Improved treatment of thermal expansion}

\begin{figure}
\begin{center}
\includegraphics[width=8.5cm,angle=0]{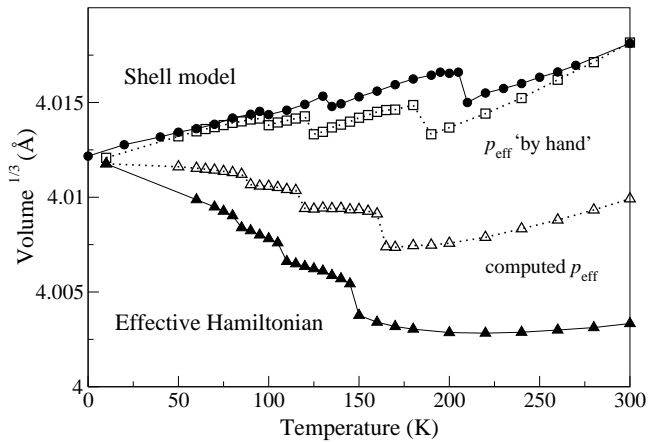}
\end{center}
\caption{Pseudocubic lattice parameter $a = V^{1/3}$
($V$ = cell volume) for BaTiO$_3$ as a function of the temperature as
predicted by the effective Hamiltonian (full triangles) and by the shell
model (full circles).
Open triangles and squares correspond to effective Hamiltonian results under an
external pressure adjusted `by hand' and  {\it ab-initio} respectively
(see text for details).}
\label{fig:vol1-3}
\end{figure}

Given the conclusion of Sec.~V.D, we are strongly motivated to improve
the effective-Hamiltonian treatment of thermal expansion.  First, we
investigate the thermal expansion in more detail.
Figure~\ref{fig:vol1-3} shows the pseudocubic lattice parameter
$a = V^{1/3}$ as a
function of temperature as predicted by the shell model (full
circles) and by the effective Hamiltonian (full triangles). Both
models exhibit volume anomalies (``kinks'') at the ferroelectric
phase transition temperatures, indicative of
their first-order character. However, the overall trends in
volume vs.~temperature are quite different.

The thermal expansion displayed by the shell model, after a
proper rescaling of temperature and cell parameter, 
closely resembles that of real BaTiO$_3$.~\cite{shi52}
This virtue of the model is related to the fact
that it includes all the degrees of freedom of the system and a
sufficiently accurate description of their relevant anharmonicities.  The
effective Hamiltonian, on the other hand, does not properly account
for the thermal expansion of the system, and actually leads to a {\it
contraction} with increasing temperature in the range of the polar
phases.  The reason for such a contraction is that the volume is
strongly coupled to the magnitude of the local dipoles and, as these
tend to {\it decrease} with increasing temperature as the paraelectric
phase is approached, the volume tends to decrease as well.
Equivalently, the thermal contraction can be attributed to negative
Gr\"uneisen parameters associated with portions of the relevant 
branches; these are overwhelmed by positive contributions from
higher modes in the shell-model system, but not in the
effective-Hamiltonian system where the higher modes are absent.

We next ask what happens if the effective-Hamiltonian simulations are
carried out with a cell volume that is constrained `by hand' to have
the correct dependence on temperature as given by the shell-model system.
A simple way of doing this in practice is to apply a (negative)
external pressure $p_{\rm eff}$ to the effective-Hamiltonian system;
this fictitious pressure can be thought of as arising from the thermal
expansion effects of the excluded modes.
We implement this approximately by taking
$p_{\rm eff}$ to be linear in temperature in such a way that the
effective-Hamiltonian equilibrium volume coincides with the shell-model
one at two temperatures, taken to
be 10 and 300\,K, bracketing the relevant range. We then find that
$p_{\rm eff}$ is required to be $-1.8$\,GPa at 300\,K while almost
vanishing at 10\,K. The results of MC simulations under this
external pressure are presented in Fig.~\ref{fig:vol1-3} (open
squares); the values of the
transition temperatures are listed in Table~\ref{tab:tc}.  The
improvement in the agreement with the shell-model calculations, in
particular in the case of the C--T transition temperature, is
remarkable.

However, such an ad-hoc approach is not consistent with the spirit
of first-principles based approaches; one would prefer
a way to calculate the effective pressure $p_{\rm eff}$ {\it
ab-initio}. We have attempted to do so by employing the so-called
quasiharmonic approximation (see Chapter~25 of Ref.~\onlinecite{ashc}). 
Within this approximation, the pressure that
develops in a harmonic crystal with volume-dependent
phonon frequencies is (see Eq.~(25.5) of Ref.~\onlinecite{ashc})
\begin{eqnarray}
p &=& - \frac{\partial}{\partial V} \left( U^{\rm eq} + \sum_{{\mathbf k}s}
\frac{1}{2} \hbar w_s({\mathbf k}) \right) \nonumber \\
&& \quad + \sum_{{\mathbf k}s} \left( 
-\frac{\partial (\hbar w_s({\mathbf k}))}{\partial V} \right)
\frac{1}{e ^{\beta \hbar w_s({\mathbf k})} - 1},
\label{eq:press}
\end{eqnarray}
where $U^{\rm eq}$ is the equilibrium energy of the system,
$\omega_{s}({\mathbf{k}})$ is the phonon frequency of branch $s$
at point $\mathbf{k}$ of the Brillouin zone (BZ), and the
summations run over all branches and $\mathbf{k}$ points. Now
the pressure exerted by the excluded modes 
is obtained from Eq.~(\ref{eq:press}) by
removing $U^{\rm eq}$ and restricting the sums to the excluded modes
$s'$.  Taking the classical limit $\hbar\rightarrow 0$
in order to compare with the classical shell model, $p_{\rm eff}$
takes the form
\begin{equation}
{\mathit{p_{\rm eff}}} = k_{B}T \frac {\partial} {\partial V} 
\left( {\displaystyle\sum_{ {\mathbf{k}} s^{'}}} 
\ln \omega_{s^{'}} ( {\mathbf{k}} ) \right)\\
\label{eq:effpres}
\end{equation}
which is linear in temperature and proportional to the
volume derivative of the phonon frequencies.

We must be cautious about the approximations involved in the use of
Eq.~(\ref{eq:effpres}) or its quantum-mechanical version. The
quasiharmonic approach is not really well suited to deal with phase
transitions, which are strongly anharmonic phenomena. Using it in the
present context relies on the assumption that the excluded modes
(more precisely, the volume derivatives of their
frequencies) are not significantly affected by the strong
fluctuations and phase transitions associated with the relevant local modes.

In order to assess the utility of the quasiharmonic approach here, we
have focused on the cubic-to-tetragonal transition and calculated
$p_{\rm eff}$ using the volume dependences of the excluded-mode frequencies
in the cubic paraelectric phase.\cite{w}  We find that the BZ sum in
Eq.~(\ref{eq:effpres}) can be evaluated with good accuracy using
information from the high-symmetry k-points only, and that the sum of
logarithms of the hard-mode frequencies depends linearly with
volume in the relevant volume range, thus allowing us to take $p_{\rm
eff}$ as independent of volume.
The $p_{\rm eff}$ calculated in this way shows improved
agreement with the exact shell-model results as regards both
the transition temperatures (denoted by ``computed
$p_{\rm eff}$'' in Table~\ref{tab:tc}) and the thermal expansion
(open triangles in Fig.~\ref{fig:vol1-3}). 
However, the results are
still far from satisfactory, with a substantial error remaining
for the C--T transition temperature.  These
discrepancies are probably connected with the shortcomings of
the quasiharmonic approximation discussed above.
Further investigations along these lines are clearly
needed, but fall beyond the scope of the present paper.

In summary, the proposed correction based on the quasiharmonic
approximation of Eq.~(\ref{eq:effpres}) accounts correctly for only a
fraction (perhaps a third) of the thermal-expansion error.
Unfortunately, then, we are not yet in a position to propose a
fully ab initio approach to the thermal expansion problem in
the context of effective-Hamiltonian methods.

\section{Conclusions}

The main weakness of our
effective-Hamiltonian description of shell-model BaTiO$_3$ is the poor
description of the thermal expansion. Focusing on the cubic to
tetragonal transition temperature ($T_{CT}$), we have found that the
effective Hamiltonian produces a 28$\%$ error, while we can estimate
from our constant-volume calculations that this error should be reduced
to 5$\%$ if the thermal expansion were properly modeled.  We have also
seen that including the thermal expansion `by hand' allows us to
reduce the error to about 12$\%$; in other words, this correction
seems to account for 70$\%$ of the total error associated with the
thermal expansion.

It is tempting to apply these same percentages in order to estimate
the sources of error arising in the comparison of the first-principles
effective Hamiltonian transition temperatures with real experiment.
However, this should be done cautiously.  For example, anharmonicities or
thermal-expansion effects might either be exaggerated or
underestimated by the shell model. With this in mind, we consider the
effective-Hamiltonian study of BaTiO$_3$ by Zhong {\it et al.}\, which led to
$T_{CT}=300$~K, 25$\%$ below the experimental value of
400\,K. This was a classical calculation; should quantum effects be
considered, the calculated $T_{CT}$ would be smaller by about 30\,K,
\cite{ini02} and thus the error would go up to $\sim$30$\%$.

We performed classical MC simulations with the effective Hamiltonian
of Zhong {\it et al}.\ including the thermal expansion of the system
`by hand' under the condition that the computed volume should coincide
with the experimental one at $T = 473$~K.  This resulted in an error
of 18$\%$ in $T_{CT}$, which would become $\sim$25$\%$ if we include
the estimated quantum effects.  Hence, it seems that the improvement in this
case is not as large as it was for the effective Hamiltonian fitted to
shell-model BaTiO$_3$. If we follow what we have learned from the case
of shell-model BaTiO$_3$ and
assume that including the thermal expansion `by hand' corrects
70$\%$ of the total thermal-expansion error, we can estimate that a
quantum first-principles effective-Hamiltonian calculation with
perfect thermal expansion would still result in a 20$\%$ underestimate
of $T_{CT}$. It seems reasonable to assume that Type~II
errors other than the thermal expansion, as well as details of the
fitting procedure, are responsible for a further 5$\%$ error in
$T_{CT}$. This suggests that a calculation free of Type~II
and Type~IV errors would yield a $T_{CT}$ that would still be about
15\% below experiment.  While this line of reasoning is tenuous, we
nevertheless believe it gives the best current estimate for the magnitude
of the error that should be attributed to the first-principles
methods used to construct the effective Hamiltonian (in particular,
the LDA).

In summary, in this work we have analyzed the errors associated with
the first-principles effective Hamiltonian method that has been
developed for the treatment of the thermodynamics of perovskite
ferroelectrics.  More specifically, by considering the
effective-Hamiltonian description of the shell-model for BaTiO$_3$ of
Tinte {\it et al.}, we have been able to isolate and study in detail
the errors intrinsic to the effective-Hamiltonian approximation
(Type~II errors). We have found that the main Type~II error is
associated with a poor description of the thermal expansion of the
system. We have discussed an easily-implemented first-principles
correction that takes into account some contributions of excluded modes.
Unfortunately, this
scheme seems to account for only about a third of the total
thermal-expansion error. More elaborate schemes (involving a more
thorough treatment of the couplings of the phonon modes to each other
and to strain and polarization) might substantially reduce the error,
but it remains for the future to explore and implement such schemes.
Finally, we have argued that in the case of the comparison of the
first-principles effective-Hamiltonian calculations on BaTiO$_3$ with
real experiment, Type~II errors do not seem to be responsible for the
entire discrepancy. Our results suggest that the
Type~I errors associated with the use of the LDA and other
first-principles technicalities may be of the same magnitude as the
thermal-expansion error.

This work was supported by the Center for Piezoelectrics by Design
(CPD) under ONR Grant N00014-01-1-0365.  Computational facilities for
the work were also provided by the CPD. J.I. acknowledges the
financial support of ONR Grant No. N0014-97-1-0048.


\begin{thebibliography}{99}
\bibitem{fn:phonon} For convenience, we use the terms ``phonons'' or 
``normal modes'' to refer to the eigenvectors of the force-constant 
matrix.
\bibitem{w} One could think of computing $p_{\rm eff}$ for each of the
phases of the system, which should then be treated separately in the
simulation.
\end{thebibliography}
\end{document}